\documentclass[final,3p,times,sort&compress]{elsarticle_for_arxiv}

\usepackage{wasysym}

\usepackage{amssymb}

\begin{document}

\begin{frontmatter}



\title{Finite difference method for the arbitrary potential in two dimensions: application to double/triple quantum dots
}


\author{Jai Seok Ahn}
\ead{jaisahn@pusan.ac.kr}

\address{Department of Physics and Research Center for Dielectrics and Advanced Matter Physics, \\
Pusan National University, Busan 609-735, Republic of Korea}

\begin{abstract}
A finite difference method (FDM) applicable to a two dimensional (2D) quantum dot was developed as a non-conventional approach to the theoretical understandings of quantum devices. This method can be applied to a realistic potential with an arbitrary shape. Using this method, the Hamiltonian in a tri-diagonal matrix could be obtained from any 2D potential, and the Hamiltonian could be diagonalized numerically for the eigenvalues. The legitimacy of this method was first checked by comparing the results with a finite round well with the analytic solutions. Two truncated harmonic wells were examined as a realistic model potential for lateral double quantum dots (DQDs) and for triple quantum dots (TQDs). The successful applications of the 2D FDM were observed with the entanglements in the DQDs. The level-splitting and anticrossing behaviors of the DQDs could be obtained by varying the distance between the dots and by introducing asymmetry in the well-depths. The 2D FDM results for linear/triangular TQDs were compared with the tight binding approximations.

\end{abstract}


\end{frontmatter}


\section{Introduction}
\label{}
The recent developments in quantum phenomena in mesoscopic systems predict many future applications of quantum devices, such as quantum information, quantum computing, next-generation logic, etc. A quantum dot with a submicron feature-size is considered as an artificial atom with a unique shell structure \cite{Reimann2002} that can be engineered artificially by manipulating a highly-mobile two-dimensional electron gas (2DEG) formed at the interface of a semiconductor heterostructure (GaAs/AlGaAs). The lateral confinement of a 2DEG is accomplished by shaping the local potential wells using gate electrodes. When two quantum dots are moved close enough to each other, they are considered as an artificial molecule that might be a candidate for a solid state quantum bit in a quantum computation \cite{Ekert1996,Loss1998,Wiel2003}. 

The theoretical understanding on the quantized bound states and the transport properties of QDs is based on the methods of quantum mechanics developed to date, such as perturbation theory with the tight-binding Anderson model \cite{Meir1993,Affleck2001,Lu2005}, variational calculations \cite{Bastard1983,Brus1984,LeGoff1992}, the {\bf k}$\cdot${\bf p} Hamiltonian method within the envelope-function approximation \cite{Sercel1990,Darnhofer1993,Burt1992}, density-functional theory \cite{Lee1998}, mode space approach \cite{Wang2004}, filter-diagonalization method \cite{Ren2008}, transmitting boundary method \cite{Lent1990,Wang1994}, numerical coupled-channel method \cite{Lin2001}, and direct diagonalization techniques in finite difference scheme \cite{Grundmann1995,Glutsch1996,Qu2003,Prabhakar2009}. 

Regarding the {\it realistic} potentials, theoretical modeling has a weakness. For example, the experimental data \cite{Heitmann1997} revealed the breaking of Kohn's theorem \cite{Kohn1961}. In particular, when it comes to closely-coupled shallow QDs, it is more challenging to employ the {\it ideal} parabolic confining potential rigorously to describe each QD: a harmonic potential requires an infinite range and height. Most theoretical methods assume an ideal and symmetric model potential and often recur to the expansions or approximations using the analytic basis functions \cite{Burkard1999,Schliemann2001}. Numerical methods are feasible alternatives and the finite difference method (FDM) can be one of the most powerful techniques for solving real quantum systems being considered recently \cite{Safarpour,Sadeghi,Li,Gong1,Deyasi,Gong2,Duque,Chow,Harrison,Harrison3rd,Hsieh,Gimenez}. This paper reports the capability of 2D FDM by examining double QDs (DQDs) and triple QDs (TQDs) with a model potential composed of truncated parabolic potential wells. This study first reviewed the 2D FDM with a single QD with round well, and examined the level-splittings and anti-crossing behaviors of DQDs. The 2D FDM and the tight binding approach are compared quantitatively in the linear TQDs and in the triangular TQDs.

\section{Theoretical model and validation}
\label{}

\subsection{The FDM in 2D}
\label{}

In the effective-mass approximation for a arbitrary $N$-electron quantum dot, the single-particle Schr\"{o}dinger equation can be given as
\begin{equation}
\left\{-\frac{\hbar^2{\vec\nabla}}{2 } \cdot \left[ \frac{\vec\nabla}{m^*({\vec r})} \right] - e \left(V^{ee} + V^b \right) + E^{xc} \right\} \psi({\vec r})=E\psi({\vec r}),
\end{equation}
where, $m^*({\vec r})$ is the electron effective mass, $V^{ee}$ is the electrostatic potential between electrons, $V^b$ is the confining barrier potential, and $E^{xc}$ is the exchange-correlation energy.  The Eq.~(1) can be solved self-consistently by solving the Poisson eq.~for $V^{ee}$ and by applying the Hartree or the local density approximation for $E^{xc}$ \cite{Perdew1981}. When a single electron is trapped within a quantum dot with a diameter of several tens of nanometers, the carrier density is very low, $\sim$ 10$^{12}$ - 10$^{13}$ /cm$^2$, and the contributions from the $V^{ee}$ and $E^{xc}$ can be neglected.

By applying a FDM to 2D regularly-spaced grid points with a grid-spacing, $\Delta$, Eq.~(1) can be approximated with a set of coupled finite difference equations,
\begin{equation}
\gamma \left( 4 \psi_{j,k} -  \psi_{j+1,k} - \psi_{j,k+1} - \psi_{j-1,k}  - \psi_{j,k-1}\right)  -e V^b_{j,k} \psi_{j,k}= E\psi_{j,k},
\end{equation}
where $\gamma = \hbar^2/{2m^*\Delta^2}$, $\psi_{j,k} = \psi(x_j, y_k)$, and $V^b_{j,k} = V^b(x_j, y_k)$.  By aligning the grid points with indices, $j$ \& $k$ (= 1, \ldots, $N$), into an one-dimensional sequence with an index $i \equiv (j-1)N + k$ (= 1, \ldots, $N^2$) \cite{Press1986}, a large but sparse Hamiltonian matrix, $H$, with non-zero elements $H_{i,i} = 4\gamma - eV^b_i$ and $H_{i+N,i} = H_{i,i+N} = H_{i+1,i} = H_{i,i+1} = -\gamma$ can be obtained.
In addition, the homogeneous domain is assumed to be surrounded by an impenetrable barrier, such that wavefunction vanishes outside, and  $H_{nN+1,nN} = H_{nN,nN+1} = 0$ for integer $n$. The Hamiltonian is a block tridiagonal matrix that can be diagonalized iteratively with the Krylov subspace method \cite{Arnoldi1951} realized using MATLAB code. The effective mass, $m^*$ = 0.067 $m_e$, was used for an electron in GaAs. A 300$\times$300 nm$^2$-area with a spatial-resolution $\Delta$ = 1 nm required $\sim$ $9\times 10^4$ grid-points. 

\subsection{Validation using a finite round well}
\label{}
First, the FDM was applied to a shallow quantum dot with a finite round well in 2D. This is a well-known pedagogical problem, of which the analytical solution is readily available, but is the most crucial step for legitimacy-checking and for testing the FDM code. 
The diameter, 2$R$, of the well was assumed to be 50 nm and the well was placed at the center of the 2D grids. The potential inside the well was set as a negative to allow bound states and the potential outside the domain to be set to zero, i.e. $V^b(r \leq R)$ = $-V_0$ and $V^b(r > R)$ = 0. The depth of the well, $V_0$, was varied within a range of 1-15 mV, and the energies and eigenfunctions of the bound states were calculated as functions of $V_0$.

The calculated bound-state-energies were plotted as functions of $V_0$ in Fig. 1(a) with symbols. As $V_0$ was increased from zero, the bound-state-energy decreased from zero and the trajectory of the energy points formed a branch of ground-state-energies, which are denoted as $E_1$. As $V_0$ was increased further, the number of bound-state-energies increased and new energy-branches emerged. For shallow wells with $V_0 < 6$ mV, only one branch appeared. For the intermediate wells with 6 mV $\leq$ $V_0$ $<$ 14 mV, three branches were found, and two of them were energy-degenerate, as indicated by $E_{2,L}$ and $E_{2,H}$; for the deeper wells with $V_0$ $\geq$ 14 mV, the number of branches becomes more than five including degenerate branches. Fig.~1(b) shows the calculated eigenfunctions for the well with $V_0$ = 15 mV with the contour plots. $U_1$, $U_{2,L}$, $U_{2,H}$, $U_{3,L}$, $U_{3,H}$, and $U_4$ are the eigenfunctions corresponding to the bound state energies, $E_1$, $E_{2,L}$, $E_{2,H}$, $E_{3,L}$, $E_{3,H}$, and $E_4$ of Fig. 1(a), respectively. The ground state wavefunction, $U_1$, is symmetrical in the angular direction. $U_{2,L}$ and $U_{2,H}$ characterize the first excited states, which are energy-degenerate but barely distinguishable just with their eigenvalues, $E_{2,L}$ and $E_{2,H}$. $U_{2,L}$ and $U_{2,H}$ have angular nodal lines, along $\sim$ $+45^{\circ}$ and its perpendicular direction $-45^{\circ}$, respectively, as shown in Fig.~1 (b). In addition, $U_{3,L}$ and $U_{3,H}$ characterize the degenerate second excited states with similar eigenvalues, $E_{3,L}$ and $E_{3,H}$. $U_{3,L}$ and $U_{3,H}$ have two nodal lines. Finally, the third excited state is non-degenerate and is characterized by the $U_4$ wavefunction and with the $E_4$ eigenvalue. $U_4$ does not have any node along the angular direction. Instead, it has one nodal line along the radial direction around the central maximum.

\begin{figure}[t]%
\begin{center}
\includegraphics*[scale=0.14]{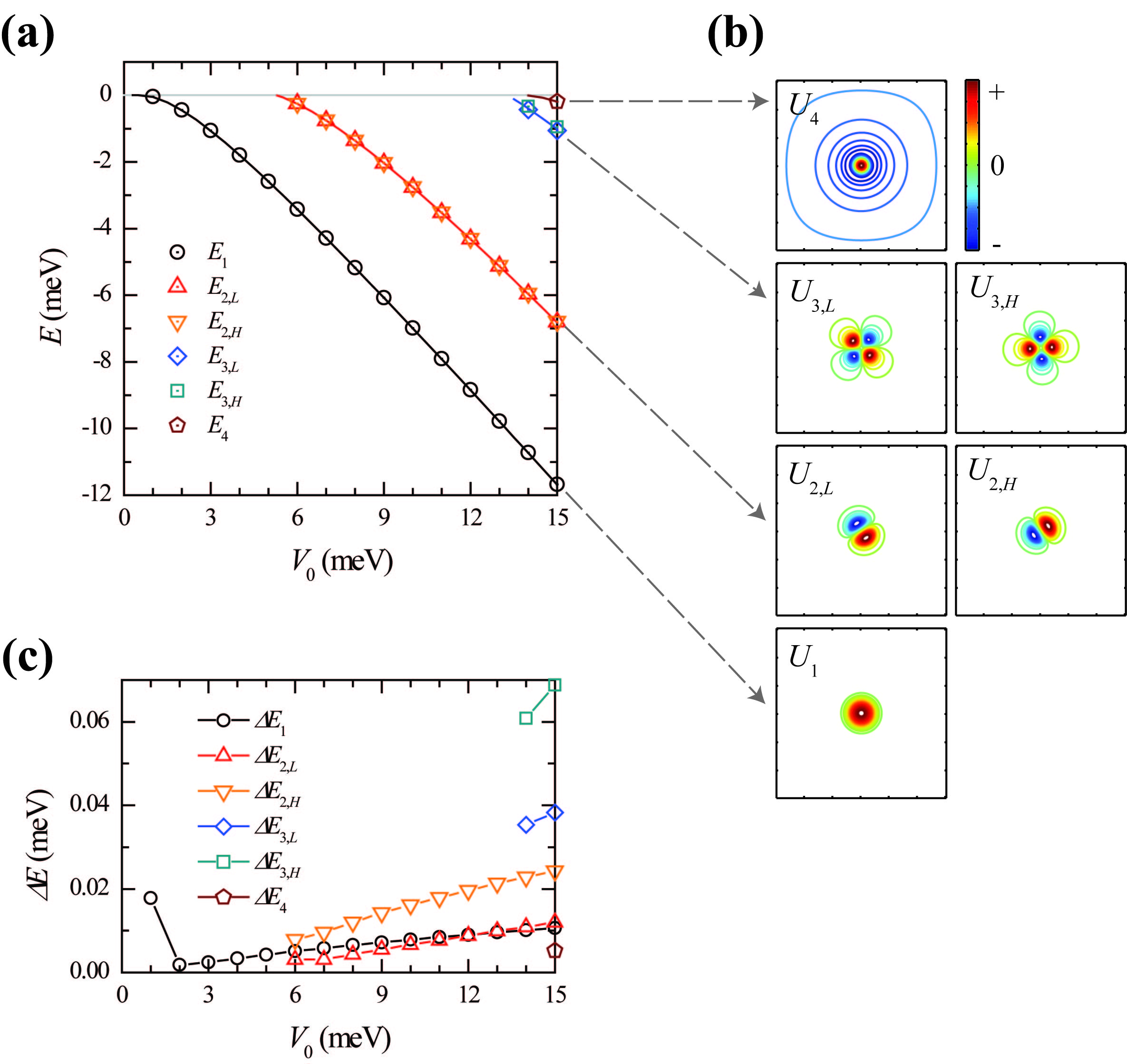}%
\end{center}
\caption[]{%
(a) Bound state energies of a shallow QD with a finite round well calculated as functions of $V_0$: $E_1$ ($\bigcirc$), $E_{2,L}$ ($\triangle$), $E_{2,H}$ ($\triangledown$), $E_{3,L}$ ($\diamond$), $E_{3,H}$ ($\square$), and $E_4$ ($\pentagon$). The results from the analytic predictions are shown with solid lines. (b) Eigenfunctions calculated for the well with $V_0$ = 15 mV. (c) Absolute energy differences between the numerical and analytic bound state energies as functions of $V_0$. 
}
\label{fig1}
\end{figure}

The calculated bound-state-energies were compared with the analytical predictions for the same problem, which is shown with four lines in Fig.~1(a), determined from the following equation, which was obtained by the continuity of the logarithmic derivative of the wavefunction at the boundary, $r$ = $R$:
\begin{equation}
\frac{kRJ_m^\prime (kR)}{J_m (kR)} = \frac{R\sqrt{2m^* e V_0 / \hbar^2 - k^2 } K_m^\prime \left( R \sqrt{2m^* e V_0 / \hbar^2 -k^2 } \right)}{K_m \left( R \sqrt{2m^* e V_0 / \hbar^2 -k^2 } \right)},               
\end{equation}
where $k$ is the wavevector in the well determined by $\sqrt{2m^* (E + e V_0) / \hbar^2}$. Here, $J_m(kr)$ is the Bessel function of the first kind and $K_m \left(r\sqrt{2m^* e V_0 / \hbar^2 -k^2 } \right)$ is the modified Bessel function of the second kind. They are proportional to the wavefunctions for the inside- and outside of the well, respectively. By solving Eq.~(3), the bound-state-energy levels can be obtained for the azimuthal quantum number, $m$'s. The lowest energy branch corresponds to the state of $m = 0$. The branches for the first and second excited states correspond to $m = \pm 1$ and $\pm 2$ states, respectively. The branch for the third excited state corresponds to the higher-momentum solution with $m = 0$. The physical meaning becomes more clearer by comparing the FDM eigenfunctions with the analytic eigenfunctions in the well, $\sim J_m (kr) e^{\pm im\phi}$. The ground state eigenfunction, $U_1$, has an asymmetric wavefunction similar to the ideal $\sim J_0 (kr)$ shape in the well. The $U_{2,L}$ or $U_{2,H}$ wavefunction for the degenerate first excited state has a $\sim J_1 (kr)\cos\phi$ or $\sim J_1 (kr)\sin\phi$ shape with the nodal line along the $\pm xy$-direction. In addition, the $U_{3,L}$ or $U_{3,H}$ wavefunction for the degenerate second excited state has a $\sim J_2 (kr)\cos2\phi$ or $\sim J_2 (kr)\sin2\phi$ shape. The $U_4$ wavefunction for the third non-degenerate excited state is interpreted as having a higher wavevector $k$ than the others, such that a radial nodal line, which is characterized by the first zero of $J_0 (kr)$, occurs within the well. Therefore, the FDM results reproduce the analytic predictions successfully for both eigenvalues and eigenfunctions. 

The absolute energy differences, $\Delta E$, between the numerical and analytic bound-state-energies, as shown in Fig.~1(c), reveal the limitation of the FDM; $\Delta E$ increases with increasing $V_0$. This effect is interpreted as a numerical artifact originating from the finite momentum. To describe the exponential decay of a wavefunction correctly, one requires an infinite number of Fourier components in principle. On the other hand, numerically, it is limited by $\sim 2\pi/\Delta$, where $\Delta$ is the grid spacing used for the FDM. Such an effect becomes more evident in $\Delta E$ with a larger $V_0$. Because the wavefunction tends to localize tightly within the well, it requires the higher momentum components. In addition, such effect is more pronounced for the degenerate excited states with non-zero $m$ values.  For $V_0$ = 15 mV, the energy-separations between the degenerate states are separated by $\sim$ 20 $\mu$eV (between $\Delta E_{2,L}$ and $\Delta E_{2,H}$) and $\sim$ 30 $\mu$eV (between $\Delta E_{3,L}$ and $\Delta E_{3,H}$), which are much larger than the absolute errors, $\lesssim$ 10 $\mu$eV, for the (non-degenerate) $m = 0$ states, $\Delta E_1$ and $\Delta E_4$. This effect can be attributed to the limited angular momentum of the FDM.

\section{Results and discussion}
\label{}

\subsection{Entanglement of the symmetric double quantum dots (DQDs)}
\label{}
To elucidate the interaction between the quantum states this section begins with double quantum dots (DQDs). To make two independent QDs interact with each other, the following three conditions need to be met: ($i$) the energy levels need to be shallow enough to have sufficient probability outside the well, ($ii$) the distance between QDs should be close enough, and ($iii$) the energy levels of each QD must be close to each other. The lateral coupling of the {\it identical} DQDs is modeled by the potential,
\begin{equation}
V^b({\vec r}) = \min \left\{0,~\frac{m^*\omega_0^2}{2e}\left[ \left( {\vec r - \vec r_1} \right)^2 - R_0^2 \right],~\frac{m^*\omega_0^2}{2e}\left[ \left( {\vec r - \vec r_2} \right)^2 - R_0^2 \right] \right\},                       
\end{equation}
which consists of two truncated harmonic wells centered at ${\vec r_1}$ and ${\vec r_2}$. Each dot with the oscillator frequency, $\omega_0$, is confined spatially within a barrier radius, $R_0$. When $|{\vec r_1} - {\vec r_2}| \leq 2R_0$, this model potential allows a coalesced snowman-shaped potential in the lateral DQD devices \cite{Blick1998,Loss2000,Chen2004,Hatao2005,Petta2005}. This model is more realistic than the previous quartic potential \cite{Burkard1999,Schliemann2001} (of one dimensional potential of fourth order polynomial for DQDs) or series coupled-DQDs \cite{Aguado2003} (with a simplification to tight binding model with only two parameters $t$ and $U$) because the interdot tunneling, $t$, distance, $d$, and (two dimensional) size, $R$, can be considered separately as illustrated in Fig.~2(a). The potential disintegrates into two separate wells in the limit $|{\vec r_1} - {\vec r_2}|  \gg 2a_0$, where $a_0 \equiv \sqrt{\hbar/m^*\omega_0}$ is the effective Bohr radius of each dot. For each dot with $R_0 = 25$ nm and a well-depth $V_0 \equiv  m^*\omega_0^2 R_0^2/2e = 5$ mV, only one bound state was permitted at the energy level of -1.160 meV and $a_0 \simeq 16$ nm. 
Note that the predicted value by using an ideal (i.e. infinite range) parabolic well, $E_{g\rm,~para} = \hbar\omega_0 - eV_0 \simeq -0.746$ meV, deviates from the calculated level, -1.160 meV. Two identical QDs were assumed to be separated spatially with the center-to-center distance, $d = |{\vec r_1} - {\vec r_2}|$, being varied from 50 to 150 nm, i.e. $\sim 3 a_0 - 10 a_0$.

\begin{figure}[t]%
\begin{center}
\includegraphics*[scale=0.13]{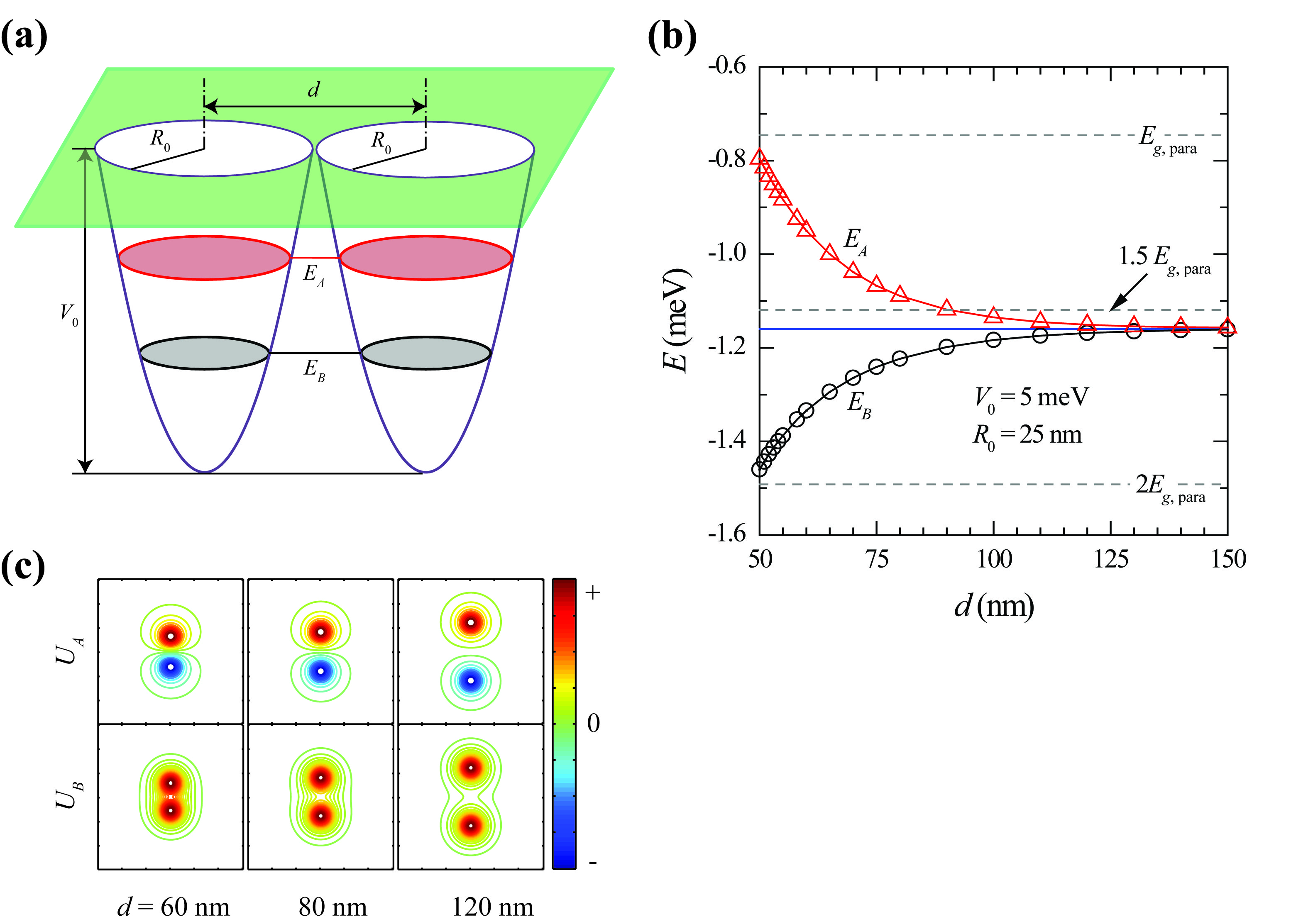}%
\end{center}
\caption[]{%
Level-splittings by entanglements in the identical DQDs. (a) A schematic diagram of the model potential in Eq.~(4), which is composed of two truncated harmonic wells. (b) Molecular bound state energies and (c) contour maps of eigenfunctions of the DQDs calculated as functions of the center-to-center distance $d$. The solid line in (b) designates the level of a single isolated QD, $\simeq$ -1.160 meV.
}
\label{fig2}
\end{figure}

The calculated energy levels and eigenfunctions show clear indications of molecular bonding for a small $d$. Fig.~2(b) shows the energies as a function of $d$. As $d$ decreases from 150 nm to 50 nm, the initially (almost) degenerate energies become separated into two different levels, $E_B$ and $E_A$, gradually. $E_B$ becomes lower and $E_A$ becomes higher than the energy level of a single QD (shown with solid line). The lower energy level, $E_B$, is interpreted as a bonding state $\sigma$ using the terms of the molecular orbital states. In contrast, the higher $E_A$ level is interpreted as an anti-bonding state, $\sigma^*$. Therefore, the energy separation between $E_B$ and $E_A$ is a measure of the entanglement in the DQDs. 
Fig.~2(c) presents the eigenfunctions, $U_B$ and $U_A$, corresponding to the $E_B$ and $E_A$ levels along with the contour maps, selectively for $d = 60$, 80, and 120 nm. The $U_B$ function has the same sign (or phase) on both centers of QDs, i.e. it is symmetric. On the other hand, the $U_A$ function shows a sign change across a nodal line, which is in conformity with the midmost line between the QDs, i.e. it is anti-symmetric. The molecular bonding can be characterized by the population of a wavefunction at the mid-zone, and becomes more covalent with decreasing $d$. These features are strongly correlated with the interaction strength between QDs, as characterized by the energy separation between the $E_B$ and $E_A$ levels. Surprisingly, the entanglement is evident even at a large distance, $\sim 8a_0 \simeq 130$ nm.

\subsection{Anticrossing in the asymmetric DQDs}
\label{}
To facilitate an interaction between QDs, the energy levels of each QD must be close to each other, but what is sufficient closeness? In addition, some applications require detuning of the energy levels between the two dots \cite{Aono2001,Cota2005}. To answer this question, two QDs with different atomic energy levels are required, and FDM is unquestionably the best suited for this purpose. The asymmetry in the potential can be introduced by detuning the radius $R$ or the frequency $\omega$ of each dot. Here, a decision was made to detune the frequency. The interaction in the {\it asymmetric} DQDs was modeled by the potential,
\begin{equation}
V^b({\vec r}) = \min \left\{0,~\frac{m^*\omega_1^2}{2e}\left[ \left( {\vec r - \vec r_1} \right)^2 - R_0^2 \right],~\frac{m^*\omega_2^2}{2e}\left[ \left( {\vec r - \vec r_2} \right)^2 - R_0^2 \right] \right\},                       
\end{equation}
which consists of two harmonic wells, QD1 and QD2, with the oscillator frequencies, $\omega_1$ and $\omega_2$, respectively, centered at ${\vec r_1}$ and ${\vec r_2}$, with $R_0 = 25$ nm as illustrated in Fig.~3(a).
$V_{0,1}$  ($\equiv  m^*\omega_1^2 R_0^2/2e$) was fixed to 5.0 mV, whereas $V_{0,2}$ ($\equiv  m^*\omega_2^2 R_0^2/2e$) was varied in the range of 0-10 mV. $d$ was also assumed constant to be 60 nm ($\simeq 4 a_0$).

\begin{figure}[t]%
\begin{center}
\includegraphics*[scale=0.14]{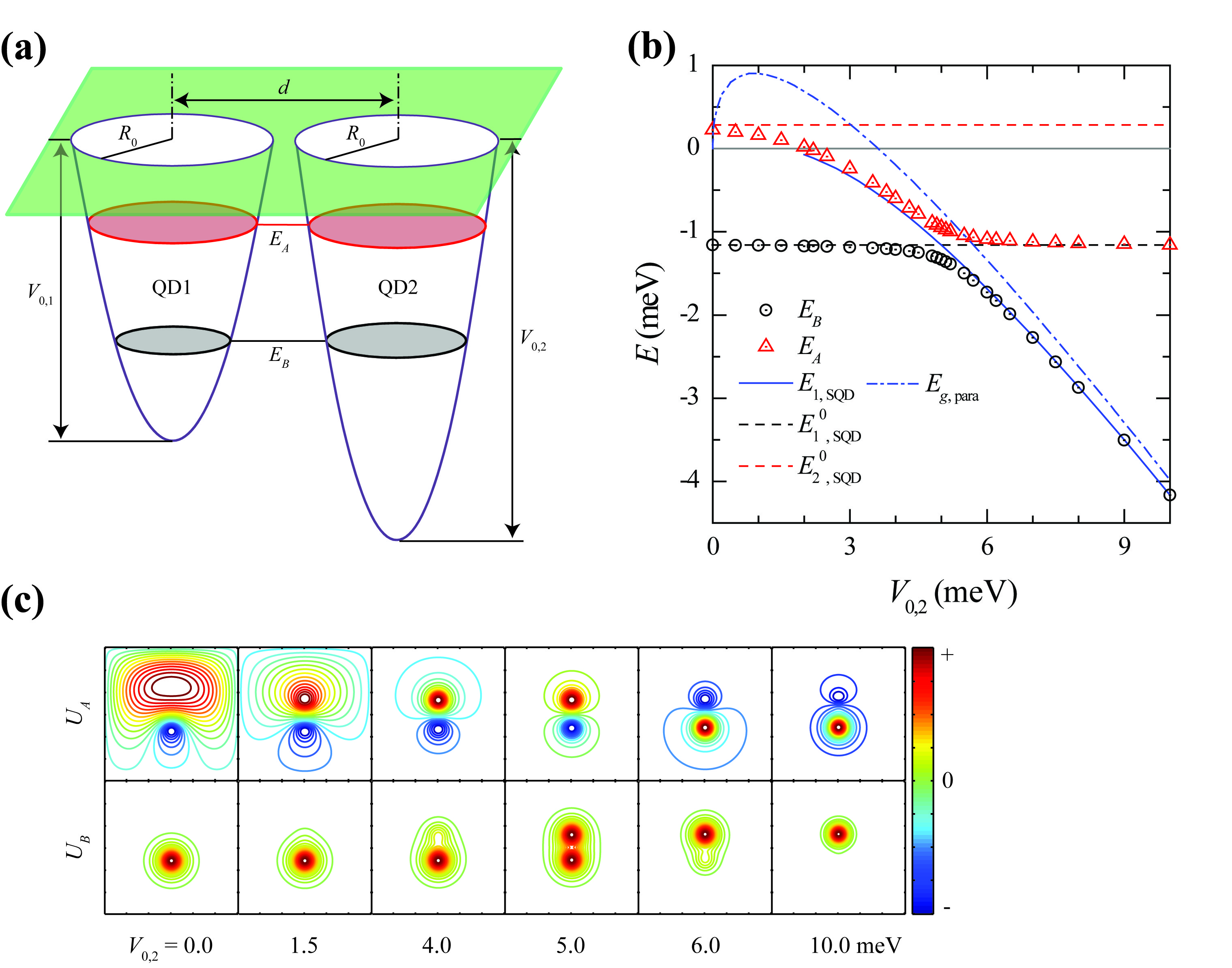}%
\end{center}
\caption[]{%
(a) A schematic illustratioin of the model potential for asymmetric DQDs in Eq.~(5). 
(b) Anticrossing of energy levels in the DQDs: bonding level $E_B$ ($\bigcirc$) and anti-bonding level $E_A$ ($\triangle$). Depth of QD1, $V_{0,1}$, was fixed to 5 mV, whereas that of QD1, $V_{0,2}$, was varied in the range of 0-10 mV. Solid line: the bound state energies of a single isolated QD. Dashed lines: the two lowest energy levels of a single isolated QD with $V_0$ = 5.0 mV: $E^0_{1,\rm SQD}$ = -1.160 and $E^0_{2,\rm SQD}$ = 0.285 meV. Dash-dotted line: the lowest energy level of an ideal parabolic-well.
(c) Contour maps of bonding ($U_B$ ) and antibonding ($U_A$) eigenfunctions of the asymmetric DQDs for $V_{0,2}$ = 0.0, 1.5, 4.0, 5.0, 6.0, and 10.0 mV.
}
\label{fig3}
\end{figure}

The calculated energy levels show a realistic view of generic anticrossing behavior \cite{Wiel2003} of the asymmetric DQDs by a tunnel-coupling.  Fig.~3(b) shows the molecular energy levels as functions of $V_{0,2}$. As stated before, the lower energy level $E_B$ can be assigned as a bonding state and the higher $E_A$ level as an anti-bonding state. For comparison, the calculated lowest energy branch of a single isolated QD ($E_{1,\rm SQD}$) was also plotted as a function of $V_{0,2}$ in a range of 2-10 mV with a solid line. The dash-dotted line shows the prediction by using an ideal (i.e. infinite range) parabolic well, $E_{g\rm,~para} = \hbar\omega_2 - eV_{0,2} = (\hbar/R_0)\sqrt{2 e V_{0,2}/m^*} - eV_{0,2}$, which deviates again from the calculated branch at the lower $V_{0,2}$ but starts to converge to it at the higher $V_{0,2}$. 
The dashed lines depict the lowest two atomic energy levels of a single isolated QD with $V_0 = 5.0$ mV, which are -1.160 ($E^0_{1,\rm SQD}$) and 0.285 ($E^0_{2,\rm SQD}$) meV. As $V_{0,2}$ increases from zero to 5.0 mV, the lowest two energy levels of DQDs form two separate branches, $E_B$ and $E_A$, which deviate from the atomic $E^0_{1,\rm SQD}$ and $E^0_{2,\rm SQD}$ levels. In particular, the $E_A$ branch rapidly follows the $E_{1,\rm SQD}$ branch. For $V_{0,2}$ in the range of 5-10 mV, the $E_B$ branch follows the $E_{1,\rm SQD}$ branch and the $E_A$ branch converges to the atomic $E^0_{1,\rm SQD}$ level. Therefore, the $E^0_{1,\rm SQD}$ level and $E_{1,\rm SQD}$ branch constitute asymptotic curves. The deviations of the $E_B$ and $E_A$ branches from the asymptotic curves are most clearly noticeable at $V_{0,2}$s within a narrow range of $\sim$ 4.5-5.5 mV, and the deviations from the atomic levels, i.e. the degree of anticrossing behaviors, can be interpreted as a measure of entanglement. This anticrossing behaviors for the tunnel-coupled DQDs can be described most simply by the quantum mechanical two-level system \cite{Wiel2003}. The molecular energy levels, $E_A$ and $E_B$, can be expressed in terms of the eigenvalues of the uncoupled double dots and the matrix element for tunneling ($t$) as $(E^0_{1,\rm SQD}+E_{1,\rm SQD})/2 \pm \sqrt{(E^0_{1,\rm SQD}-E_{1,\rm SQD})^2/4 + |t|^2}$.

The eigenfunctions, $U_B$ and $U_A$, corresponding to the $E_B$ and $E_A$ branches, respectively, were plotted with contour maps in Fig.~3(c), selectively for $V_{0,2}$ = 0.0, 1.5, 4.0, 5.0, 6.0, and 10.0 mV. The $U_B$ has the same sign (or phase) on the entire domain but $U_A$ shows a sign change across a nodal curve between the QDs. The pair of perfect symmetric and anti-symmetric wavefunctions, i.e. the duo of anticrossing levels, can be found only when the depths of the two QDs are equal, i.e. $V_{0,2}$ = 5.0 mV. The symmetric point can have a unique description as a static picture for the coherent charge oscillations observed in the charge qubit systems \cite{Hayashi2003,Gorman2005,Koppens2006}. Therefore when time-evolution is allowed from the point, this FDM can be extended further to examine the coherent (adiabatic) dynamics of such systems with spontaneous symmetry breaking. It is possible within FDM using a unitary time-evolution operator in the Crank-Nicolson algorithm \cite{CrankNicolson}.  As asymmetry is introduced in the potentials, the molecular wavefunctions tend to localize one of the QD sites because the ground state settles at the deepest potential well. When the $V_{0,2}$ is smaller (or shallower) than $V_{0,1}$ (= 5.0 mV), the bonding state $U_B$ is relatively confined to the QD1 site, whereas the anti-bonding state $U_A$ is localized to the QD2 site. In addition, when the $V_{0,2}$ is larger (or deeper) than the $V_{0,1}$, the $U_B$ and  $U_A$ moves to the QD2-site and QD1-site, respectively. 

\subsection{Linear triple quantum dots (LTQDs)}
\label{}
By placing three identical QDs at regular intervals along a straight line, we examined the linear triple QDs (LTQDs) \cite{Hsieh}.
For each QD with truncated shallow harmonic well as in the previous sections, it was assumed that $V_0$ = 5 mV and $R_0$ = 25 nm.
Three separate bound levels were calculated using the FDM.
Figure 4(a) shows the energy levels as functions of the interdot distance $d$.

\begin{figure}[t]%
\begin{center}
\includegraphics*[scale=0.45]{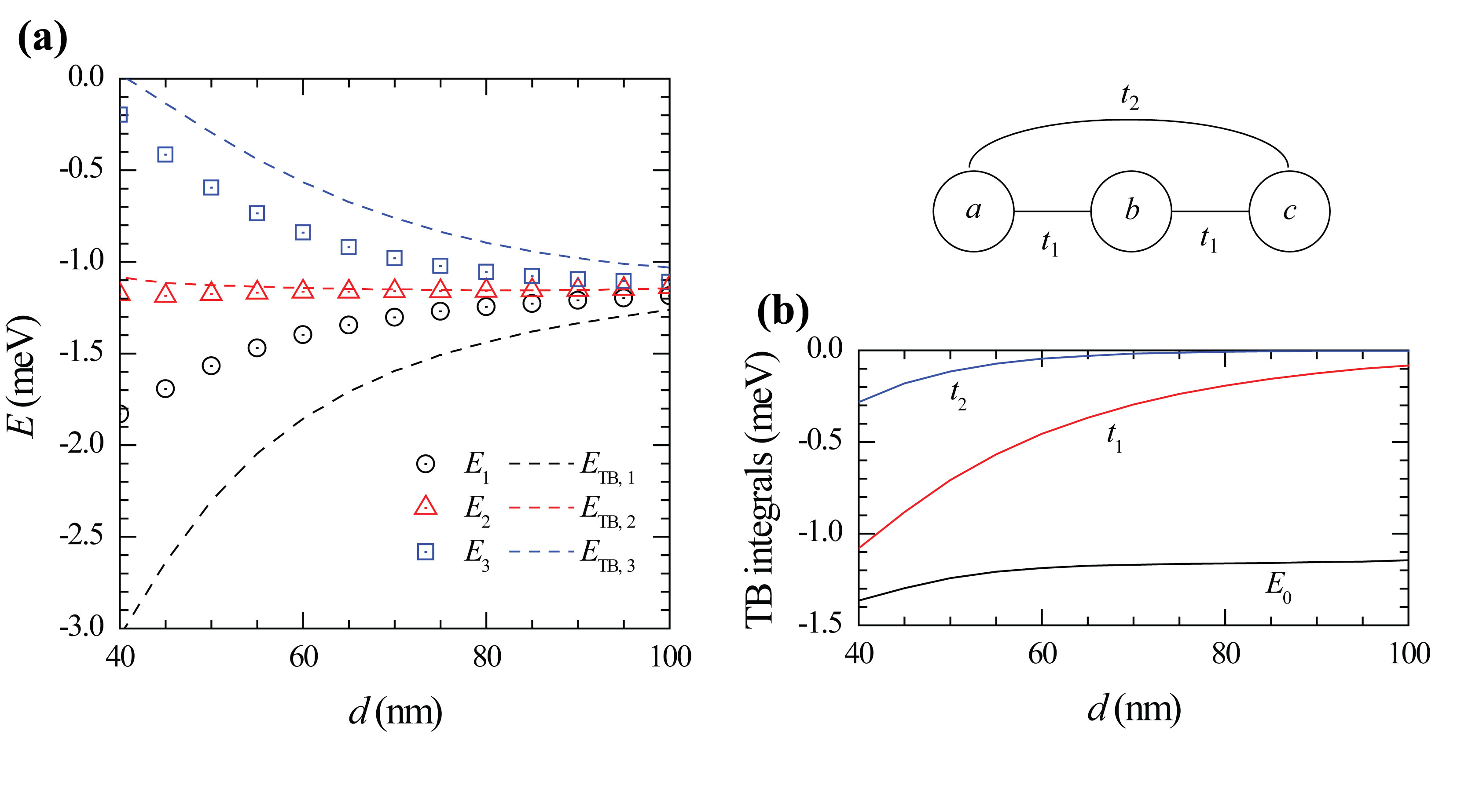}%
\end{center}
\caption[]{%
(a) Bound levels of LTQDs as functions of distance $d$: FDM results (symbols) and TB model (lines)
(b) TB integrals for the model.
Schematic diagram: overlap integrals, $t_1$ and $t_2$, in LTQDs.
}
\label{fig4}
\end{figure}

It is intriguing to compare the FDM results with the tight-binding (TB) model calculations. 
When we approximate the LTQDs by using a TB model with three atomic basis wavefunctions, the TB Hamiltonian becomes
\begin{equation}
H_{TB} = \left[
\begin{array}{ccc}
E_0 & t_1 & t_2\\
t_1 & E_0 & t_1\\
t_2 & t_1 & E_0 
\end{array} \right]~.
\end{equation}
The diagonal element, $E_0$, defines the atomic binding energy evaluated as $<a|H|a>$. 
The off-diagonal matrix elements describe the electron overlap (or tunneling) between dots.
The integrals, $t_1$ = $<$$a|H|b$$>$ and  $t_2$ = $<$$a|H|c$$>$ are for nearest neighbor and for next-nearest neighbor, respectively.
Here, $|a$$>$, $|b$$>$, and $|c$$>$ represent the atomic eigenfunctions centered on each dot.
We calculated the three TB parameters ($E_0$, $t_1$, and $t_2$) as shown in Fig. 4(b), by using the FDM Hamiltonian for LTQDs and the FDM eigenfunction for an isolated QD.
The three TB binding energies were evaluated as $E_0 + \frac{t_2}{2} \pm \sqrt{2 t_1^2 + (\frac{t_2}{2})^2} $ and $E_0 - t_2$.
The TB approximations reproduced the FDM results qualitatively as shown in Fig. 4(a), but their accuracy was limited.

\subsection{Triangular triple quantum dots (TTQDs)}
\label{}
If you put in the same separation between three identical QDs in a two-dimensional plane, it becomes the triple QDs with an equilateral triangular shape \cite{Gimenez,Hsieh}. 
The energy levels of triangular triple QDs (TTQDs) as a function of interdot distance $d$ were also examined using the FDM.  
When the Hamiltonian for the TTQDs were diagonalized using the FDM, two separate energy branches were obtained. As shown in Fig. 5(a), the upper branch is doubly-degenerated.

The TB Hamiltonian for the TTQDs is
\begin{equation}
H_{TB} = \left[
\begin{array}{ccc}
E_0 & t & t\\
t & E_0 & t\\
t & t & E_0 
\end{array} \right]~,
\end{equation}
where, $t$ = $<$$a|H|b$$>$ =$<$$b|H|c$$>$ = $<$$c|H|a$$>$ and  $E_0$ = $<$$a|H|a$$>$ represents the nearest neighbor electron tunneling and the atomic energy level. 
The TB binding energies were evaluated as $E_0 + 2 t$ and $E_0 - t$ (degenerated).
When we write the eigenfunctions using the linear combination of atomic orbitals (LCAO), the ground state is approximated as $\frac{|a> + |b> + |c>}{\sqrt{3}}$ and the (energy-degenerate) excited states as $\frac{|a> - |b>}{\sqrt{2}}$ and $\frac{|a> - |c>}{\sqrt{2}}$.
The discrete (three-fold) rotation symmetry is reflected from the degenerate-excited states. 
The double-degeneracy is connected to the 2D nature of an electron in TTQDs, and the degeneracy can be removed by introducing an asymmetry between QDs or by applying a magnetic field perpendicular to the plane.

\begin{figure}[t]%
\begin{center}
\includegraphics*[scale=0.45]{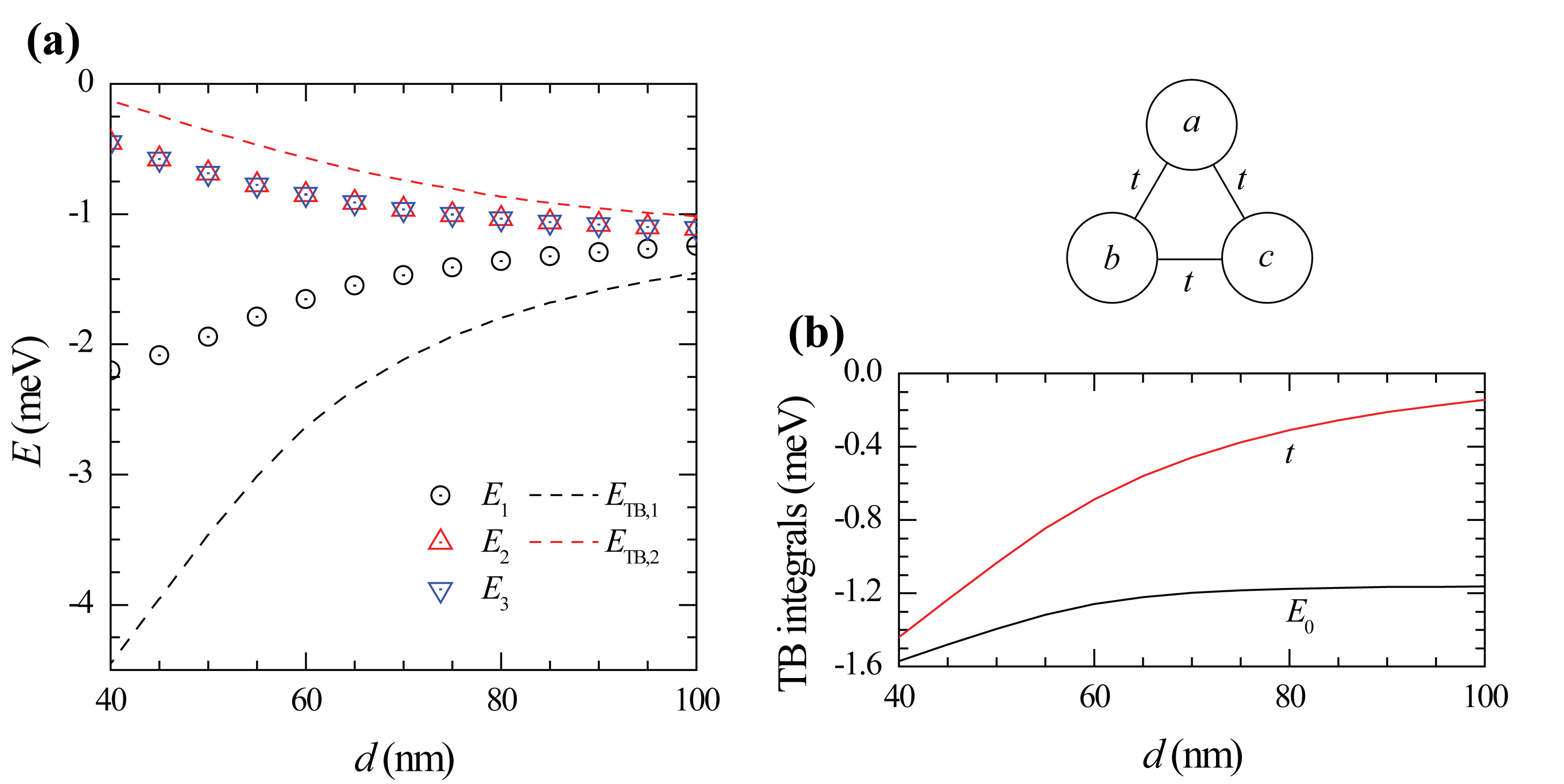}%
\end{center}
\caption[]{%
(a) Bound levels of TTQDs as functions of distance $d$: FDM results (symbols) and TBB model (lines). (b) TB integrals for the model. Schematic diagram: overlap integral $t$ in TTQDs.
}
\label{fig5}
\end{figure}

\section{Conclusions}
\label{}
A 2D FDM applicable to {\it realistic} quantum devices with a 2D potential of arbitrary shape was developed. 
The results showed that the Hamiltonian for the device can be described in a tri-diagonal matrix regardless of the model potential, and can be diagonalized numerically for the eigenvalues and eigenfunctions. The developed method was tested quantitatively using a well-known finite round well problem. The small numerical artifacts could be analyzed as the finite size effect of linear/angular momentum. The successful applications of the 2D FDM, as a powerful technique in solving a real quantum system, were demonstrated with the DQDs and TQDs. The entanglements in lateral DQDs were modeled by a model potential with double truncated parabolic potential wells, which allows independent considerations of the interdot tunneling, interdot distance, and dot-size. The level-splittings and anticrossing behaviors of the DQDs could be obtained quantitatively with high-precision. The quantitative differences were disclosed between the 2D FDM results and the TB calculations of LTQDs and TTQDs. 

\section*{Acknowledgment}
\label{}
This work was supported for two years by Pusan National U. Research Grant. 








\end{document}